\documentclass[]{spie}  

\usepackage{aasmacros}
\usepackage{amsmath,amsfonts,amssymb}
\usepackage{graphicx}
\usepackage[colorlinks=true, allcolors=blue]{hyperref}

\newcommand{\micron}{$\mu$m}
\newcommand{\spherex}{SPHEREx}

\newdimen\sa  \newdimen\sb
\def\arcsec{\ifmmode {^{\scriptstyle\prime\prime}}
          \else $^{\scriptstyle\prime\prime}$\fi}
          
\def\arcmin{\ifmmode {^{\scriptstyle\prime}}
          \else $^{\scriptstyle\prime}$\fi}     
          
\def\parcs{\sa=.07em \sb=.03em
     \ifmmode \hbox{\rlap{.}}^{\scriptstyle\prime\kern -\sb\prime}\hbox{\kern -\sa}
     \else \rlap{.}$^{\scriptstyle\prime\kern -\sb\prime}$\kern -\sa\fi}

\def\deg{\ifmmode^\circ\else$^\circ$\fi}

\title{SPHEREx: NASA’s Near-Infrared Spectrophotometric All-Sky Survey}

\author[a]{Brendan P. Crill}
\author[a]{Michael Werner}
\author[b]{Rachel Akeson}
\author[c]{Matthew Ashby}
\author[d]{Lindsey Bleem}
\author[a,e]{James J. Bock}
\author[f]{Sean Bryan}
\author[e]{Jill Burnham}
\author[h]{Joyce Byun}
\author[a]{Tzu-Ching Chang}
\author[i]{Yi-Kuan Chiang}
\author[e]{Walter Cook}
\author[g]{Asantha Cooray} 
\author[e]{Andrew Davis}
\author[a]{Olivier Dor{\'e}}
\author[a]{C. Darren Dowell}
\author[b]{Gregory Dubois-Felsmann}
\author[h]{Tim Eifler}
\author[b]{Andreas Faisst}
\author[d]{Salman Habib}
\author[e]{Chen Heinrich}
\author[d]{Katrin Heitmann}
\author[e]{Grigory Heaton}
\author[i]{Christopher Hirata}
\author[e]{Viktor Hristov}
\author[e]{Howard Hui}
\author[j]{Woong-Seob Jeong}
\author[e]{Jae Hwan Kang}
\author[e]{Branislav Kecman}
\author[b]{J. Davy Kirkpatrick} 
\author[e]{Phillip M. Korngut}
\author[h]{Elisabeth Krause}
\author[b,j]{Bomee Lee}
\author[k]{Carey Lisse}
\author[b]{Daniel Masters}
\author[f]{Philip Mauskopf}
\author[c]{Gary Melnick}
\author[e]{Hiromasa Miyasaka}
\author[g]{Hooshang Nayyeri} 
\author[a]{Hien Nguyen}
\author[c]{Karin {\"O}berg}
\author[e]{Steve Padin}
\author[b]{Roberta Paladini}
\author[g]{Milad Pourrahmani}
\author[j]{Jeonghyun Pyo}
\author[e]{Roger Smith}
\author[j]{Yong-Seon Song}
\author[l]{Teresa Symons}
\author[b]{Harry Teplitz}
\author[c]{Volker Tolls}
\author[a]{Stephen Unwin}
\author[f]{Rogier Windhorst}  
\author[j]{Yujin Yang}
\author[l]{Michael Zemcov}

\affil[a]{Jet Propulsion Laboratory, California Institute of Technology, 4800 Oak Grove, Pasadena CA 91109, USA}
\affil[b]{IPAC, Caltech, 770 S. Wilson Ave, Pasadena, CA 91125, USA}
\affil[c]{Center for Astrophysics $|$ Harvard \& Smithsonian, 60 Garden St., Cambridge, MA 02138, USA}
\affil[d]{Argonne National Laboratory, High-Energy Physics Division, 9700 S. Cass Avenue, Argonne, IL 60439,USA}
\affil[e]{California Institute of Technology, 1200 E. California Blvd, Pasadena, CA 91125, USA}
\affil[f]{Arizona State University, Department of Physics, P.O. Box 871504, Tempe, AZ 85287, USA}
\affil[g]{University of California Irvine, 4186 Frederick Reines Hall, Irvine, CA 92697, USA}
\affil[h]{Steward Observatory, The University of Arizona, Tucson, AZ 85721, USA}
\affil[i]{Department of Astronomy, The Ohio State University, 140 W. 18th Avenue, Columbus, OH 43210 USA}
\affil[j]{Korea Astronomy and Space Science Institute, Daejeon, 34055, Korea}
\affil[k]{JHU-APL, SES/SRE, Bldg 200/E206, 11100 Johns Hopkins Road, Laurel, MD 20723, USA}
\affil[l]{Rochester Institute of Technology, College of Science, 74 Lomb Memorial Drive, Rochester, NY 14623,USA}

\authorinfo{Further author information:\\B.P.C.: E-mail: {\tt bcrill@jpl.nasa.gov}, Telephone: +1 818 354 5416}

\begin{document} 
\maketitle

\begin{abstract}
\spherex, the Spectro-Photometer for the History of the Universe, Epoch of Reionization, and ices Explorer, is a NASA MIDEX mission planned for launch in 2024.  \spherex\ will carry out the first all-sky spectral survey at wavelengths between 0.75\micron\ and 5\micron\  with spectral resolving power $\sim$40 between 0.75 and 3.8\micron\  and $\sim$120 between 3.8 and 5\micron\   At the end of its two-year mission, \spherex\ will provide 0.75-to-5\micron\ spectra of each 6\parcs2$\times$6\parcs2 pixel on the sky -- 14 billion spectra in all.   This paper updates an earlier description of \spherex\ presenting changes made during the mission's Preliminary Design Phase, including a discussion of instrument integration and test flow and a summary of the data processing, analysis, and distribution plans. 
\end{abstract}

\keywords{All-sky survey, linear variable filters, near infrared, cosmology, biogenic ices, extragalactic background light}

\section{INTRODUCTION}
\label{sec:intro}  
\spherex\ is a Medium Explorer (MIDEX) mission to undertake an all-sky spectrophotometric survey in the near-IR that will enable new investigations into highly compelling science questions of today. \spherex\ was selected by NASA to enter Preliminary Design Phase (Phase B) in 2019, and is working towards a 2024 launch.

We start this paper with a brief overview of \spherex’s scientific rationale, the hardware configuration, and the mission design, which have previously been presented in an earlier SPIE proceeding\cite{Korngut2018} (hereafter Paper 1). We go on to describe several updates to the \spherex\ system which grew out of the just-completed mission Phase B.  We discuss important features of \spherex: payload integration and testing, data processing, and science products, which were not covered in the earlier paper.  A final section presents \spherex\ in the context of other current and forthcoming space- and ground-based missions, from the point of view both of schedule and of measurement capability.  Details concerning the \spherex\ mission are available in Paper 1 and at \url{https://spherex.caltech.edu}.  \spherex\ science is well-described in three community workshop reports\cite{Dore2014,Dore2016,Dore2018}. 
\subsection{Science Themes}
The \spherex\ science team has optimized the mission to address three scientific questions consistent with the three major themes of NASA’s astrophysics program.  In particular, \spherex\ will:  
\begin{enumerate}
    \item Probe the origin of the Universe by constraining the physics of inflation, the superluminal expansion of the Universe that took place $\sim$10$^{-32}$ s after the Big Bang, by measuring galaxy redshifts over a large cosmological volume.
    \item Chart the history of galaxy formation through deep images of the continuously visible regions near the ecliptic poles, which will permit a precise determination of the spectrum of fluctuations in the extragalactic background light.    
    \item Investigate the origin of water and biogenic molecules in the early phases of planetary system formation – from molecular clouds to young stellar systems with planet-forming disks – by measuring absorption spectra to determine the abundance and composition of interstellar and circumstellar ices.

\end{enumerate}
	
As described in subsequent sections, the \spherex\ science team is committed to providing data products and scientific papers which address each of the three themes above.   However, as compelling as these three themes are, they do not encompass the full range of \spherex\ science.  The history of all-sky surveys, dating back to the pioneering photographic sky surveys of the 1940’s, makes us confident that the scientific community will use the \spherex\ data in novel ways to uncover phenomena and explore questions which we cannot imagine.   
\spherex\ maps 1.4 trillion spectral-spatial elements (voxels) over the entire sky in each of four independent surveys.  As an indication of the expected richness of the \spherex\ data set, the data are expected to include:
\begin{itemize}
    \item $>$1 billion galaxy spectra
    \item $>$100 million high-quality galaxy redshifts
    \item $>$100 million stellar spectra
    \item $>$100 thousand ice absorption spectra
    \item $>$1 million quasar spectra
    \item $>$10 thousand asteroid spectra
\end{itemize}

The \spherex\ mission will provide all images and tools to access this expansive data via a public archive hosted at NASA's Infrared Science Archive (IRSA)\footnote{\url{http://irsa.ipac.caltech.edu}}.

\section{SPHEREX PAYLOAD}

\spherex\ performs its spectroscopic survey with no moving parts in the instrument. Instead, a wide-field off-axis all-aluminum 20~cm diameter telescope feeds two 1$\times$3 mosaics of H2RG HgCdTe detector arrays\cite{Blank2012} in the focal plane separated by a dichroic beamsplitter. Linear variable filters (LVFs) fixed just above the arrays allow \spherex\  to measure spectra of the entire sky in roughly 6~months through a series of exposures using a sequence of spacecraft pointings.  The optical system has very high throughput, with each detector having a field of view of 3.5\deg$\times$3.5\deg, so that each 1$\times$3 mosaic has a field of view of 3.5\deg$\times$11.3\deg.  The shorter wavelength mosaic, which views the light reflected from the beamsplitter, contains 3 arrays with 2.6\micron\ long-wavelength.  The 3 arrays with 5.3\micron\  cutoffs, are mounted so as to detect light transmitted through the dichroic.  The parameters of the LVFs over these arrays are listed in Table~\ref{tab:spherex_bands}.  The three short wavelength arrays (and, separately, the three long wavelength arrays) are packaged together in a mechanical assembly called a focal plane assembly (FPA).

\begin{table}[ht]
\caption{\spherex\ parameters at a glance.}
\label{tab:spherex_summary}
\begin{center}
    \begin{tabular}{ll}
    \textbf{Telescope} \\
    \hline
    Telescope design & free-form three mirror anastigmat\\
    Focal ratio & f/3\\
    Effective aperture & 20~cm\\
    Field-of-view & 3.5\deg $\times$ 11.3\deg\  ($\times$2)\\
    \'{E}tendue & 1.2 m$^2$ deg$^2$\\
    Wavelength Coverage & 0.75 -- 5.0~$\mu$m\\
    Operating temperature & $<$80~K \\
    \\
    \textbf{Focal Plane} \\
    \hline
    Detectors & 3$\times$ H2RG (2.6~\micron\ cutoff) $<$80~K\\
              & 3$\times$ H2RG (5.3~\micron\ cutoff) $<$55~K\\
    Pixel count & 25M \\
    Read mode & Sample-up-the-ramp\cite{Zemcov2016}\\
    Read time & 1.5~s (nominal)\\
    Exposure time & 112.5~s (nominal)\\
    Pixel resolution & 6\parcs2 $\times$ 6\parcs2\\
    Bandpass Filters & Linear Variable Filters (See Table \ref{tab:spherex_bands})\\
    Spectral resolution ($\lambda$/$\Delta\lambda$) & 41 -- 130 (See Table~\ref{tab:spherex_bands})\\
    Point Source Sensitivity & $>$18.4 ABMag (5$\sigma$) \\ & per spectral bin (R=41) at 2\micron \\
    & (See Fig.~\ref{fig:spherex_sensitivity})\\
    \\
    \textbf{Mission} \\
    \hline
    Duration & 2 years \\
    Orbit & 700 km Low Earth \\
          & 6am ascending node \\ 
          & sun-synchronous\\
    Data rate & 120 Gbit/day \\
    Sky coverage & full sky, 4 full surveys \\
    Sky coverage - deep survey & 200 deg$^2$ \\     

    \end{tabular}
\end{center}
\end{table}

\begin{table}[ht]
\caption{Parameters of \spherex\ bands.  The minimum and maximum wavelength of each band includes padding of a quarter of a spectral resolution element to include overlap between neighboring bands and the dichroic.} 
\label{tab:spherex_bands}
\begin{center}       
\begin{tabular}{|c|c|c|c|} 
\hline
\rule[-1ex]{0pt}{3.5ex}  Band & $\lambda_{\mathrm{min}}$  [\micron]& $\lambda_{\mathrm{max}}$  [\micron] & R ($\lambda$/$\Delta\lambda$) \\
\hline
\rule[-1ex]{0pt}{3.5ex}  1 & 0.75 & 1.12 & 41   \\
\hline 
\rule[-1ex]{0pt}{3.5ex}  2 & 1.10 & 1.65 & 41   \\
\hline 
\rule[-1ex]{0pt}{3.5ex}  3 & 1.63 & 2.44 & 41   \\
\hline 
\rule[-1ex]{0pt}{3.5ex}  4 & 2.40 & 3.85 & 35   \\
\hline 
\rule[-1ex]{0pt}{3.5ex}  5 & 3.81 & 4.43 & 110   \\
\hline 
\rule[-1ex]{0pt}{3.5ex}  6 & 4.41 & 5.01 & 130   \\
\hline 
\end{tabular}
\end{center}
\end{table}

\begin{figure} [ht]
   \begin{center}
   \begin{tabular}{c} 
   \includegraphics[width=\columnwidth]{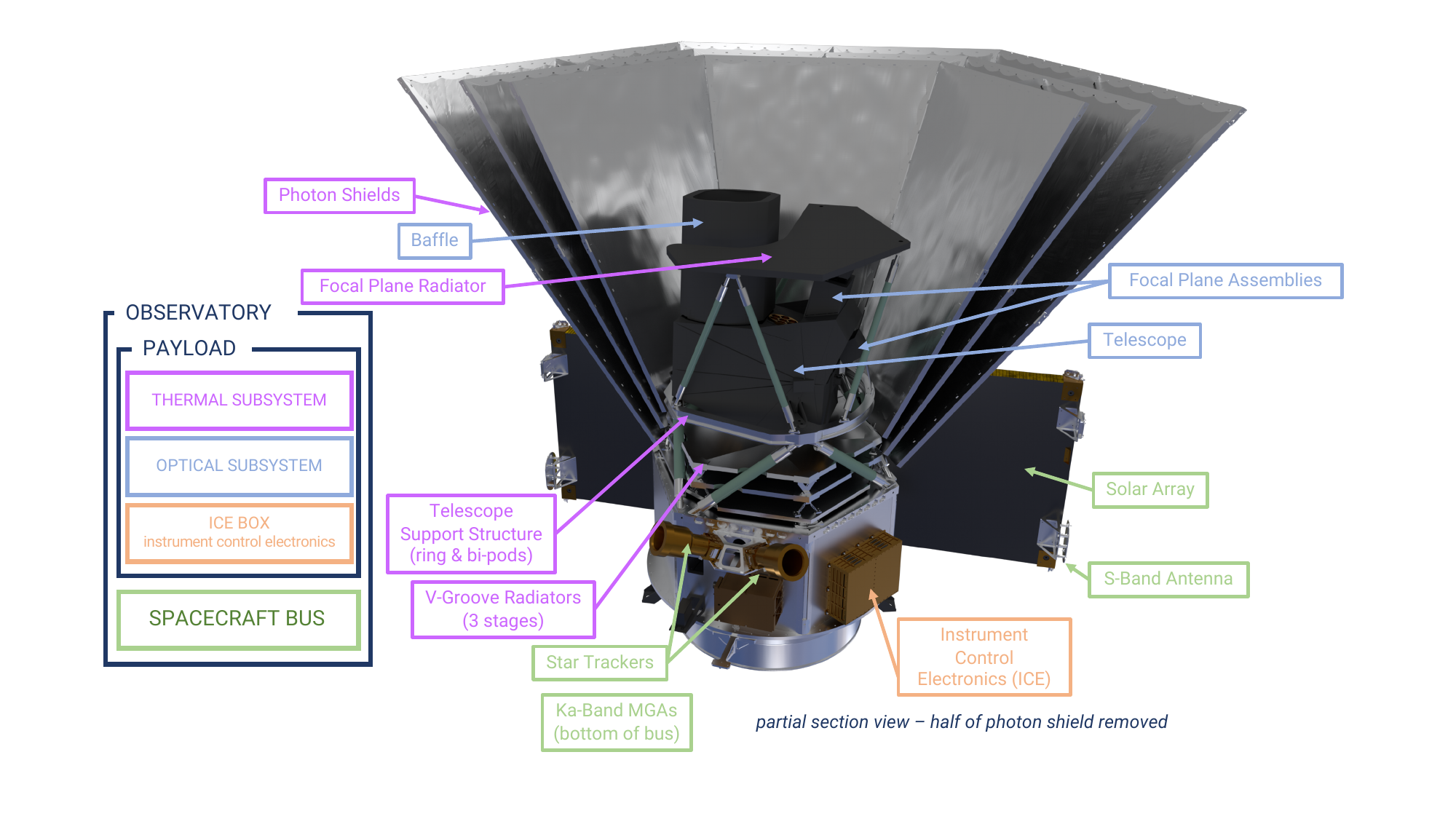}
   \end{tabular}
   \end{center}
   \caption[example] 
   { \label{fig:spherex_observatory} 
The \spherex\ observatory (photon shields cut away) showing the bus with the payload optical and thermal subsystems. The optics and detector cool passively by radiating heat to space with a 3-stage V-groove passive cooler. Photon shields protect the cooler and optics from radiation from the Sun and Earth. 
}
   \end{figure}

The telescope is passively cooled to below 80~K in low-Earth orbit by three nested V-groove radiators. An additional radiator cools the long wavelength focal plane temperature below 60~K to reduce detector dark current. Figure~\ref{fig:spherex_observatory} shows the current \spherex\ system configuration, as presented at the  Preliminary Design Review in October 2020.  

\begin{figure} [ht]
   \begin{center}
   \begin{tabular}{c} 
   \includegraphics[height=9cm]{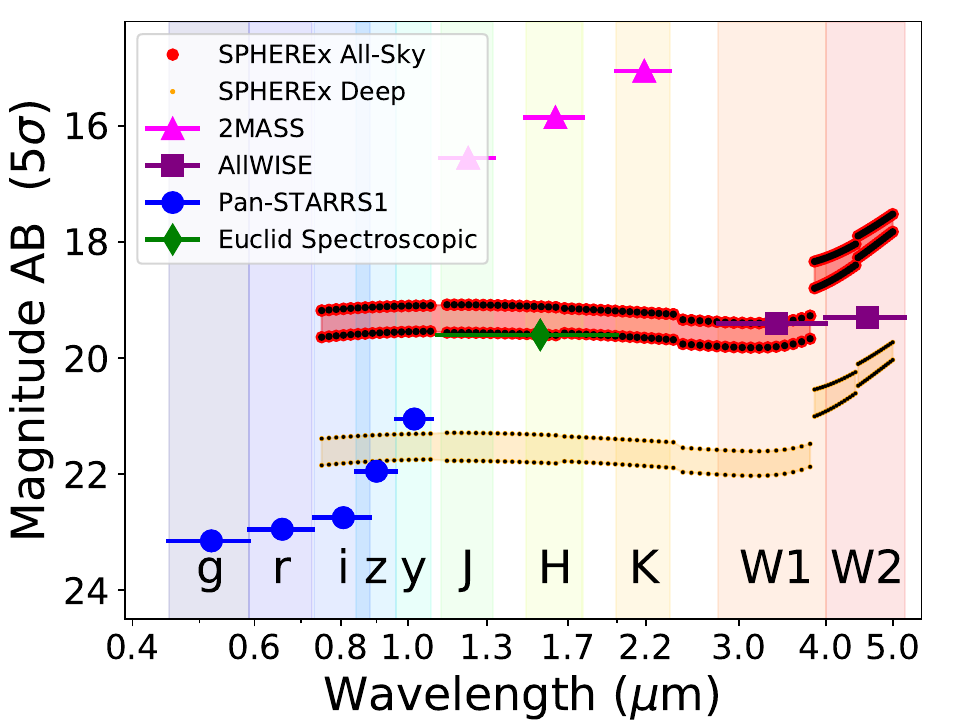}
   \end{tabular}
   \end{center}
   \caption[example] 
   { \label{fig:spherex_sensitivity} 
Sensitivity of \spherex\ and current surveys, all at 5-sigma. The \spherex\ sensitivity is quoted per spectral resolution element at the completion of the four surveys which comprise the two-year mission.  The bottom red curve corresponds to the current best estimate for the sensitivity over the entire sky, while the top red curve shows the sensitivity achieved if each subsystem just meets its performance requirements.  The orange curves show the analogous sensitivity estimates for the deep survey regions at the ecliptic poles.  These curves do not include the effects of astrophysical source confusion, which is significant at the deep survey depth.}
   \end{figure}

A yardstick with which to assess \spherex’s sensitivity can be obtained from previous missions.  In particular, 2MASS and WISE are all-sky broadband photometric surveys, each of which cataloged hundreds of millions of objects, which encompass much of the spectral range covered by \spherex.  As shown in Figure~\ref{fig:spherex_sensitivity}, \spherex\ will measure the spectrum of every object in the 2MASS point source catalog (1.2~\micron, 1.6~\micron, 2.2~\micron) to at least (40$\sigma$, 60$\sigma$, 150$\sigma$) per spectral channel.  \spherex\ will also measure spectra of essentially all objects in the WISE catalog at 3.3 and 4.8\micron, with the faintest detected at $\sim$3$\sigma$ in each spectral channel.

\section{MISSION DESIGN}

\spherex\ has two Focal Plane Assemblies (FPAs) separated by a dichroic to deliver full short- (0.75--2.44~$\mu$m) and long- (2.40--5.01~$\mu$m) wavelength coverage. Each of the six detector arrays has its own LVF. The central wavelength transmitted by each LVF varies with position, defining the passband of each detector pixel. We observe from a main pointing list that consists of telescope pointings spaced to enable \spherex\ to deliver all-sky coverage in each of our 102 spectral resolution elements. \spherex\ operates in a 700-km altitude sun-synchronous polar orbit, maintaining a 91 degree Solar avoidance angle, and less than 35 degree local zenith angle. The orbits are divided into a number of segments approximately 10 minutes in duration.  For the start of each segment, our survey planning software selects a target from the main target list based on the available pointing zone due to sun and earth avoidance, orbit data, and the coverage achieved thus far in the mission. The spacecraft slews to this target and the telescope stably points for approximately 112 seconds, obtaining a single exposure of sources at wavelengths which depend on where on the focal plane they fall.  After each exposure, the telescope is moved to the next target in the main list (approximately 12\arcmin\ away to sample the next spectral channel of the LVF), and each source is measured at a new wavelength. This is repeated for up to 4 total exposures per orbit segment, at which point the Earth pointing constraints require that the telescope undertake a larger slew of about 60\deg\ to another patch of sky, where the same process is repeated at the next orbit segment. 

The data are stored on board and downlinked several times per day as the telescope passes over the ground stations of NASA’s Near-Earth Network.  In approximately six months, as the sun-synchronous orbit tracks the Earth’s rotation around the Sun, spectra of every point on the sky are recorded in this manner.  Our baseline mission design calls for four complete surveys, which are offset from one another by half a spectral resolution element so as to achieve Nyquist sampling of the spectrum.

This survey strategy, which crosses the Galactic plane twice per orbit and revisits the poles of the orbit repeatedly, matches \spherex’s scientific objectives.  Each of the four surveys will contain Deep Fields - $\sim$100 sq degree regions near the ecliptic poles, visible on each orbit. Our main target list includes these deep fields, accumulating much deeper coverage than the all-sky survey by taking advantage of the frequent revisits due to our orbit.  The Deep Fields will be used to probe the infrared background by assembling  multi-wavelength spatial anisotropy maps in a search for the integrated light of the earliest galaxies.  When \spherex\ crosses the Galactic plane, interstellar dust and gas obscures its view of the extragalactic sky, but the large numbers of stars at these low Galactic latitudes make this an ideal region for the study of interstellar and circumstellar ices.  At intermediate latitudes \spherex\ will carry out the extensive extragalactic surveys required to study cosmic inflation.

\section{SPACECRAFT}
Ball Aerospace will deliver the \spherex\ spacecraft and telescope.  The \spherex\ mission leverages the modularity and heritage of the Ball Configurable Platform (BCP) spacecraft.  The flexible BCP architecture readily meets \spherex\ requirements.  The \spherex\ spacecraft is the latest of the Ball BCP product line, which has been flight demonstrated in a similar configuration on multiple programs, including WISE, launched in 2009 and still operational.  \spherex\ will also benefit from Ball’s providing the spacecraft for the IXPE X-Ray mission, planned for launch in 2021, which precedes \spherex\ in NASA’s Explorer program.  WISE, IXPE and other recent Ball spacecraft developed for low orbit operations have requirements broadly similar to those of \spherex.

During the recently completed Phase B study, the \spherex-specific requirements on the spacecraft have been defined and satisfied by modelling and analysis.  Extensive study has shown that the spacecraft attitude control system can meet \spherex\ pointing accuracy (34\arcsec\ one sigma per axis) and stability (1\parcs4 one sigma per axis) requirements while retaining the agility required to execute the long slews and short steps demanded by the mission profile described above.  

\subsection{Design Changes in Phase B}
As one would expect, the recently completed Phase B study led to maturation in system design, as well as updates, in comparison to that presented in Paper 1.   Two updates merit specific attention:  Changing from deployable to fixed photon shields, and replacing the SIDECAR ASICS, previously baselined for the detector array readout, with an alternative design referred to as Video-8.

The adoption of fixed photon shields eliminates the risk of on-orbit deployment, assures the required reflecting surface, and minimizes the possible, and difficult-to-estimate, impact of the somewhat floppy deployed photon shields on the performance of \spherex’s pointing and control system.  The fixed photon shields fit nicely in the generic rocket fairing baselined for the \spherex\ launch, even as the length of the shields has been increased by 195~mm to reduce Earthshine on a  telescope baffle that has been lengthened to improve rejection of lunar stray light.  The fixed shields will be made of honeycomb sandwich panel frames, carbon fiber face sheet and aluminum core.  The shields are covered with vacuum-deposited Aluminum-coated Kapton blankets on each side of each panel.  The outermost blanket has Teflon coating to reduce degradation due to atomic oxygen at the 700~km orbit.

The instrument concept presented in Paper 1 used cryogenic SIDECAR ASICs for array readout, amplification and A/D conversion, with the digital data being transferred to an instrument electronics package on the spacecraft for Level-0 processing, compression, and conversion to images for downlink. In the Phase B implementation of \spherex, the functionality of the SIDECARs has been replaced by a readout system designed at Caltech around a custom-built ASIC chip referred to as the Video8.  Unlike the SIDECARs, these chips operate at room temperature.  A single chip reads eight fully differential detector channels, provides box filtering and “sample and hold” outputs to external A/D converters. 
To meet the challenging correlated noise requirements imposed by our galaxy formation science theme, the \spherex\ Video8 readout design is tailored to our needs.  This is accomplished using a combination of techniques that seek to minimize spurious clustering signals in the photocurrent image data.  The approach is threefold, and is enumerated below.
\begin{enumerate}
\item We control intrinsic sources of 1/f noise in hardware extensively.  This includes providing a very stable detector bias voltage from a thermally regulated and heavily conditioned power supply.  We also use an amplifier chain with well characterized and minimal temperature coefficients.
\item Embed referencing throughout the readout frame, with extra clock cycles included in which the inputs to the Video8 are shorted.  We refer to these as “phantom pixels”, and they provide a real time, low read noise measurement of the 1/f drifts that can be subtracted in analysis.
\item Choose a multiplexing scheme that maps residual 1/f noise to spatial frequencies outside the band of interest.  We refer to this technique as “row chopping”, and it measures sequential rows of physically adjacent pixels at temporally separated intervals throughout a frame.
\end{enumerate}
We have fabricated the flight Video8 chips and demonstrated the above techniques in a prototype electronics system operating in concert with a cold H2RG read out integrated circuit (ROIC).  With this suite of processes, we have demonstrated noise that follows a white spatial power spectrum down to a multipole of 500 assuming \spherex’s plate scale of 6\arcsec per pixel.

\section{Instrument Integration and Performance Testing}

The \spherex\ instrument (Fig.~\ref{fig:spherex_observatory}) has three major subsystems: Optical, including the telescope, the cryogenic dichroic beam splitter and the focal plane assemblies; Thermal, including the V-groove radiators, the photon shields, and the focal plane radiator which cools the long wavelength focal plane assembly; and Electronic, which includes the VIDEO-8 devices for operating the detectors, FPGAs which execute the sample up the ramp computations, associated bias and housekeeping circuitry, instrument control software, and ambient temperature and cryogenic cabling.  

Each subsystem is assembled and undergoes appropriate performance and environmental testing at the subsystem level.  Importantly, however, the small size of \spherex\ allows us to test the assembled payload, without the photon shields but including the other components of the thermal subsystem,   under realistic thermal and radiative conditions.

\begin{figure} [ht]
   \begin{center}
   \begin{tabular}{c} 
   \includegraphics[width=\columnwidth]{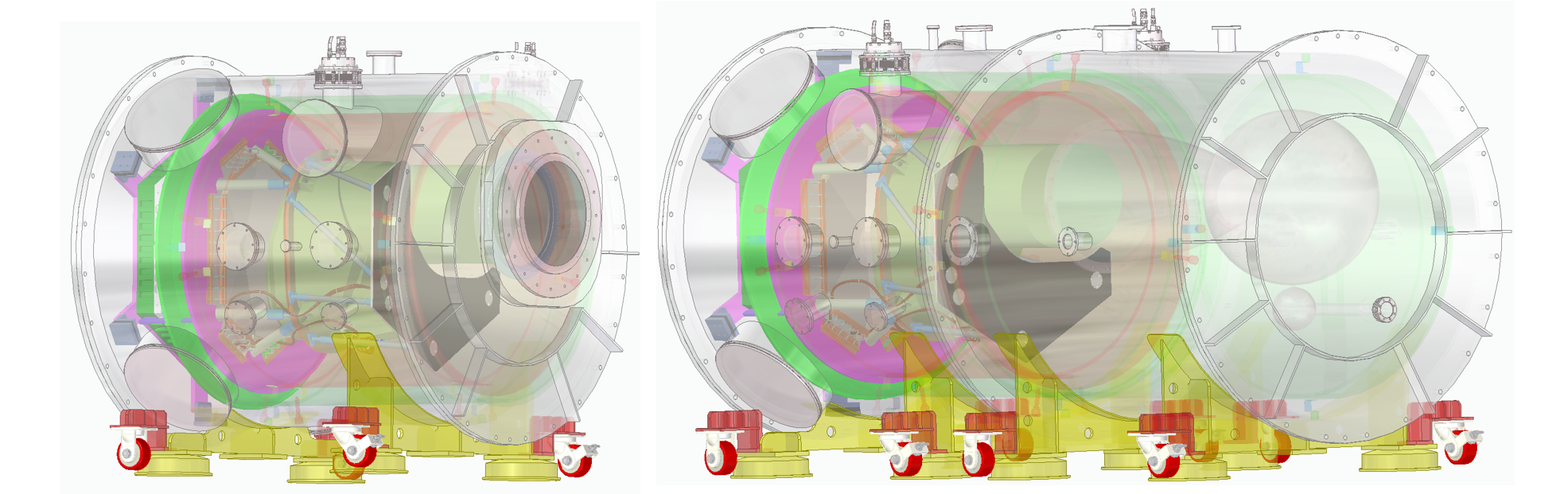}
   \end{tabular}
   \end{center}
   \caption[example] 
   { \label{fig:test_chamber} 
\spherex\ instrument cryogenic test chamber. The left panel shows the chamber in optical mode and the right shows the dark mode. }
   \end{figure}

The test plan is based on a dedicated Ground Support Equipment (GSE) cryogenic chamber, developed by the Korea Astronomy and Space Science Institute (KASI) and delivered to Caltech. Its design follows that of previous KASI chambers for end-to-end system testing for  the NISS (Near-infrared Imaging Spectrometer for Star formation history) space mission\cite{Jeong2018}.  KASI fabricates and tests the performance of the chamber prior to delivery. Once at Caltech, the chamber undergoes several cryogenic validation runs prior to the optical tests.  The \spherex\ chamber (Fig.~\ref{fig:test_chamber}) cools the telescope and detectors with two 2-stage pulse tube coolers with 80~K and 20~K stages.  The 80~K cooler stages connect to the outer V-groove and provide an outer radiation shield; the 20 K cooler stages connect to the telescope, FPA radiator, and inner radiation shield.  We note that the performance of the observatory-level radiative cooling system, including the photon shields, is completely tested in a separate facility.

The chamber supports two test modes: optical mode with an optical window, and dark mode with a cryogenic integrating sphere. Optical mode (Fig.~\ref{fig:test_chamber}, left panel) uses a large sapphire vacuum window and a cryogenic cold filter to control thermal emission from the laboratory, so that collimated infrared light can shine into the chamber to measure focus. Note that the thermal filter can have low optical efficiency for these tests. The all-reflecting collimator uses a pinhole source, scanned parallel to the optical axis to determine best focus over the full FOV. 

Dark mode configuration (Fig.~\ref{fig:test_chamber}, right panel) removes the windows and installs a cold integrating sphere with a $\sim$60cm Winston cone aperture in the forward bulkhead. The sphere is an all-aluminum design with a sandblasted, gold-coated inner surface, fiber coupled to optical sources in the laboratory via a small integrating sphere illuminated through a small window. Illuminating the sphere with external spectral sources allows us to measure the full spectral response of every pixel in the focal plane. We can close the optical port with a cold shutter to give a low photon background for measuring end-to-end dark noise performance, with the detectors, instrument electronics, and cryogenic harnesses in their flight configuration.  The chamber also includes a a cryogenic shutter mechanism to control the ambient light levels in this configuration.

The chamber tests begin with the first thermal vacuum test (TVAC-1) in optical mode to characterize the end-to-end cryogenic focus, and check that the FPAs are co-registered to the telescope to within $\pm$10\micron. Based on these results, we adjust the FPA mount at room temperature, and repeat the cold focus test in TVAC-2. We then vibration test the thermal and optical subsystems at room-temperature (as appropriate for an ambient-temperature launch). After vibration, TVAC-3 tests optical focus to confirm repeatability. Finally, the chamber switches to dark mode for TVAC-4 to measure the dark noise performance, characterizing the full spectral response of every pixel in the focal plane and obtaining a preliminary flat-field response and calibration. Of these measurements, only the spectral response is a required ground-based data product. The final calibration and flat field response come exclusively from flight data; there will be on average 60,000 SNR$>$10 non-saturating stars in each 3.5\deg$\times$3.5\deg\ exposure, minimizing the complexity and schedule risk of the ground- based test program.

\section{DATA PRODUCTS}
\begin{figure} [ht]
   \begin{center}
   \begin{tabular}{c} 
   \includegraphics[width=\columnwidth]{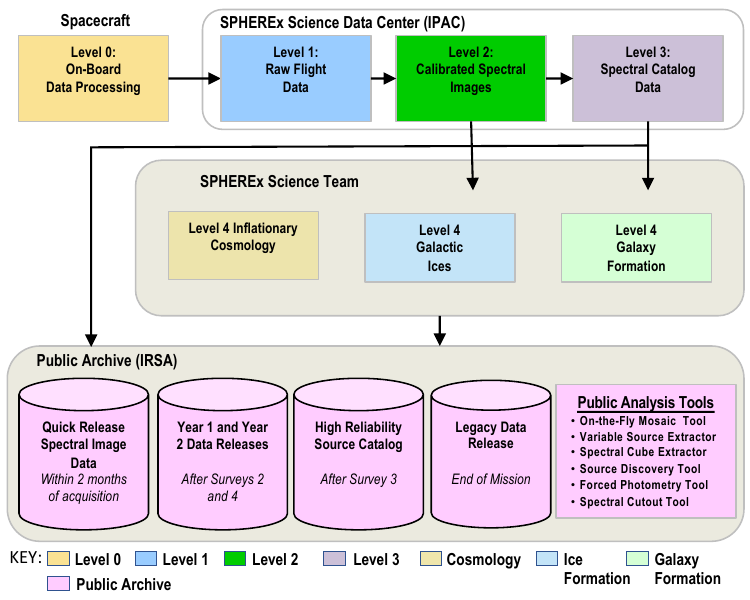}
   \end{tabular}
   \end{center}
   \caption[example] 
   { \label{fig:data_flow} 
\spherex\ data processing flow. }
   \end{figure}

During its two-year mission, \spherex\ collects a full-sky survey of 1.4 trillion spatial-spectral measurements at 6\parcs2 resolution.  The \spherex\ Science Data Center (SSDC), located at IPAC at Caltech, handles the main Level 1-3 data processing pipeline and data releases, leveraging extensive experience with infrared space data processing. The \spherex\ Science Team (SST) processes the data at Level 4, as summarized in Figure~\ref{fig:data_flow}. In addition, the SST partners with IPAC to produce the algorithms used for the Level 1-to-3 processing described here, but IPAC is responsible for converting the algorithms to robust code for use during the mission.  The SST also provides the SSDC with the Reference Catalog, a catalog of known source positions to carry out photometric measurements needed for the Inflation and Ices investigations.
The data products flow from on-board data reduction to scientific catalogs as follows:

{\bf Level-0 On-Board Data}: The on-board electronics process the detector data, producing best-fit sample up the ramp (SUR) slope estimates and quality flags, and compress the science data, which are telemetered to the ground at a rate of $\sim$120 Gbit per day. The JPL Mission Operations System (MOS) depacketizes the data and transfers the compressed photocurrent images, spacecraft attitude data, and other ancillary flight data to the SSDC.

The on-board SUR routine\cite{Zemcov2016} flags pixels for cosmic rays, non-linearity, and saturation.  
Based on available WISE and HST data, we estimate that cosmic rays will trigger a data flag in $\sim$0.22\% of the pixels in a typical $\sim$120~s observation. 
While \spherex\ will continue observations while the spacecraft is in the South Atlantic Anomaly (SAA), the images are excluded from the science investigations. 
Though \spherex\ observations still cover the full sky by scheduling observations when the observatory is outside of the SAA.  
At $\sim$30\deg\ Galactic latitude at a wavelength of 2\micron, sources brighter than $\sim$ 11 AB mag trigger a saturation flag on $<$0.1\% of pixels, and the SUR algorithm provides a linear signal estimate using only samples when the integrated charge is less than a set value. 
Sources brighter than $\sim$6 AB mag saturate in $<$2 samples so that $<$0.001\% of pixels are cut without a slope estimate.
From simulations, we expect to flag an additional 0.13\% of pixels from hysteretic response from bright sources. 
We eliminate data which trip any of these flags from the analysis of the \spherex\ core science themes but provide estimated signals where possible for public data releases.

{\bf Level-1 Photocurrent Images in Engineering Units}: SSDC executes an automated pipeline that decompresses the flight data, synchronizes with housekeeping and spacecraft attitude data, reformats the data into FITS format images, and performs data quality assessment.  The L1 data set is in electrons per second, and comprises 42 GB/day in volume.  

{\bf Level-2 Spectral Images}: SSDC passes the image data through an automated pipeline using information from ancillary catalogs, ground calibration products, and in-flight calibration products to produce a database of calibrated and astrometrically-registered (to the Gaia reference frame) spectral images.  The PSFs within the images are also reconstructed and included as ancillary data with each Spectral Image. These calibrated Spectral Images are the basis for the Galaxy Formation theme and for the Quick Release deliveries to the public data archive at IRSA.  These images are the embodiment of the flight data taken through the LVFs as described above.

{\bf Level-3 All-Sky Spectral Catalog}: SSDC generates spectra for sources in the Reference Catalog, a selection by the science team from a number of pre-existing catalogs that will include galaxies and stellar sources.  For each source, we obtain from all the images in which that source appears a spectrum with a forced optimally-weighted photometry algorithm that uses the PSF to estimate the source flux in each observation.  This produces a catalog of fully-sampled spectra which serves as the basis of the Cosmic Inflation and Interstellar Ices science analysis. These two themes each use the galaxies or stars in the Reference Catalog, respectively, and the pipeline carries out forced photometry at the target positions to produce the \spherex\ spectrum. A subset of the sources in the L3 catalog which robustly pass data consistency checks between the first two sky surveys are chosen for a High Reliability Source Catalog (HRSC).

{\bf Level-4 Science Products} At Level 4, the processing divides among the three science themes, each of which is the responsibility of a subset of the \spherex\ science team.  The Cosmic Inflation group produces a L4 Galaxy Redshift Catalog used to compute the 3-dimensional galaxy clustering power spectrum and bi-spectrum, and inferred cosmological parameters.  The Interstellar Ices team takes sources from the spectral catalog and ancillary information to compute the underlying source spectra and the ice column densities and relative abundances.  These data make up the L4 Ice Catalog. The History of Galaxy Formation group produces L4 Deep Field Mosaic Images in broad spectral bands from the two highly redundant deep survey regions near the ecliptic poles.  They estimate the correlated noise, mask point sources from the images, compute auto- and cross-power spectra, and compare the results to a range of models for the extragalactic Background Light.

IRSA will serve the \spherex\ data in a public archive where it will be fully accessible in the context of data from other missions.  We will release the calibrated Spectral Images as Quick Release Spectral Image Data within two months of acquisition, on a continuing basis, starting 2 months after launch.  The archive will facilitate fast lookup of all images for a given point on the sky (each of which will contain the given point measured at a different wavelength).  The metadata and software tools required to convert this series of images into a spectrum will be made available as part of the release.  It will be triggered by the input of a source position, as described above for the Level 4 Galaxy Redshift and Ices Catalogs, which will produce a spectrum of the input position by forced photometry. Calibration and processing of these images is refined throughout the mission, but these releases allow immediate use of the \spherex\ data by the astronomical community.  

Table~\ref{tab:data_release} details the planned data release schedule throughout and following the completion of the \spherex\ mission.    Following each year of the mission, IRSA will release the Spectral Images reprocessed with uniform processing and the best-to-date calibration.  The first such comprehensive data release contains spectral images from the first and second full sky surveys.  We reprocess and release previous surveys in subsequent data releases as in-flight calibration products and the control of systematic errors improves during the mission.  The Year 1 and Year 2 data releases also include Data Cube products consisting of 102 single-color full sky mosaics assembled from the Spectral Image data.  Each data release includes full documentation in the form of an explanatory supplement, including code examples for reconstructing full spectra for a given sky location.  

In addition, we release a High Reliability Source Catalog (HRSC) of spectra, following the completion of the first three surveys to ensure at least two full spectral samples of every source. The catalog uses source positions from external catalogs, with highly statistically significant and reliable \spherex\ data, based on consistency between the two \spherex\ surveys, naturally rejecting spurious data, variable sources, and moving solar system objects. The astronomical community can mine the HRSC for interesting sources for further study with NASA facilities such as JWST.

The final Legacy Data Release consists of the output of the Level 4 processing from the three science themes, released at the end of the mission, after the completion of Survey 4, when we have fully quantified the data reliability, including the L4 Galaxy Redshift Catalog, Ice Catalog, and Deep-Field Mosaics.  

At each data release, SSDC transfers data to the IRSA archive, where it is made available alongside the extensive holdings from previous missions. The standard IRSA interface to the spectral image data and catalogs, as well as the database of metadata, allow visualization, search, and retrieval of \spherex\ data.  Several new software tools delivered to IRSA by SST and SSDC will enhance interaction specifically with the \spherex\ data:
\begin{itemize}
    \item \textbf{Forced Photometry} to measure spectra at user-supplied positions from the database of Spectral Images.
    \item \textbf{Source Discovery} to detect signals in user-supplied spatial regions.
    \item \textbf{On-the Fly-Mosaic} to create single-wavelength images from the Spectral Image data.
    \item \textbf{Spectral Cutout} to collect user-specified regions from the database of Spectral Images.
    \item \textbf{Spectral Data Cube Extractor} to extract a user-defined subset data cube from the full sky Data Cube.
    \item \textbf{Variable Source Inspector} to analyze \spherex\ catalog or user-derived spectra at multiple epochs.
\end{itemize}

\begin{table}[ht]
\caption{Data release schedule.} 
\label{tab:data_release}
\begin{center}       
\begin{tabular}{|p{60mm}|p{25mm}|p{75mm}|} 
\hline
\rule[-1ex]{0pt}{3.5ex}  {\bf Data Product} & {\bf Date} & {\bf Note} \\
\hline
\rule[-1ex]{0pt}{3.5ex}  Quick Release Spectral Image Data &  Launch+3 months until End of Mission & Regular releases beginning two months after survey start.  Data released within two months of acquisition   \\
\hline 
\rule[-1ex]{0pt}{3.5ex}  Spectral Images & Launch+1.5~years, Launch+2.5~years &Year 1 and Year 2 releases 
(6 months after end of surveys 2 and 4) \\
\hline 
\rule[-1ex]{0pt}{3.5ex}  All-Sky Data Cube & Launch+1.5~years, Launch+2.5~years &Year 1 and Year 2 releases \\
\hline 
\rule[-1ex]{0pt}{3.5ex}  High Reliability Source Catalog & Launch+2~years & 8 months after end of survey 3  \\
\hline 
\rule[-1ex]{0pt}{3.5ex}  Deep Field Mosaics & Launch+3~years &Legacy Data Release (12 months after end of operations)   \\
\hline 
\rule[-1ex]{0pt}{3.5ex}  Stellar Type / Ice Column Density Catalog & Launch+3~years & Legacy Data Release  \\
\hline 
\rule[-1ex]{0pt}{3.5ex}  Galaxy Redshift Catalog & Launch+3~years & Legacy Data Release  \\
\hline 
\end{tabular}
\end{center}
\end{table}

\subsection{Data Simulation}

The \spherex\ team has developed a software system called the \spherex\ Sky Simulator that generates simulated \spherex\ Spectral Image and catalog data. The simulated data is used to assist in developing software modules, to test and validate the pipeline, and to evaluate effects of systematic errors on science results.   The simulation software is meant to create  representations of the eventual mission data obtained by \spherex\ on-orbit, including realistic instrumental behavior.  

The Sky Simulator is flexible and modular, and can inject arbitrary compact and diffuse sources.  Currently the input stellar catalog includes sources matched between Gaia\cite{GAIADR2} and CatWISE2020\cite{Eisenhardt2020} in order to generate a coarse SED of the stars across the \spherex\ bands.  A zodiacal dust model based on a modification of the COBE model\cite{Kelsall1998}, incorporating the Solar spectrum\footnote{\href{https://www.nrel.gov/grid/solar-resource/spectra-astm-e490.html}{2000 ASTM Standard Extraterrestrial Spectrum Reference E-490-00}} as well as updated to more recent scattered-light measurements\cite{Tsumura2010,Tsumura2013_zodi,Matsumoto2015,Kawara2017} and Planck measurements of zodi emission\cite{planck2013-zodi}.   A model of diffuse Galactic light based on Planck and IRAS measurements of dust\cite{planck_dust} scaled to the near-infrared\cite{Lillie1976,Draine2003,Zubko2004,Tsumura2013_DGL,Arai2015} can also be included.  Diffuse sources (for example, resolved galaxies such as M31) can also be injected.  Photon noise from all simulated sky signal sources is consistently generated.   Additional signal sources are prioritized for addition to the sky can be added as needed by the project. 

\begin{figure} [ht]
   \begin{center}
   \begin{tabular}{rr} 
   \includegraphics[height=8cm]{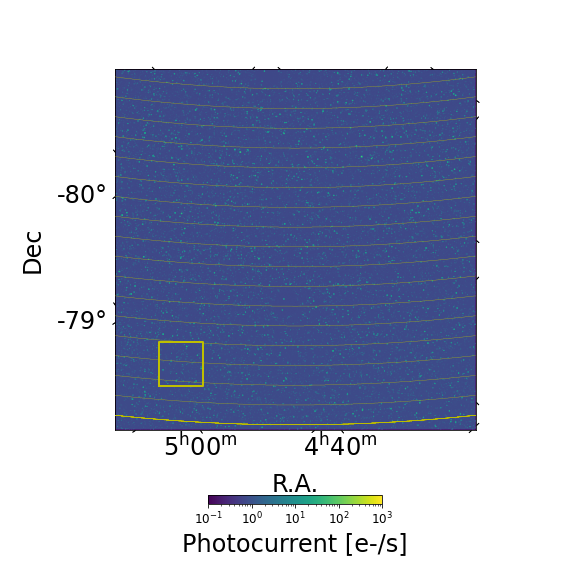} &
  \includegraphics[height=8cm]{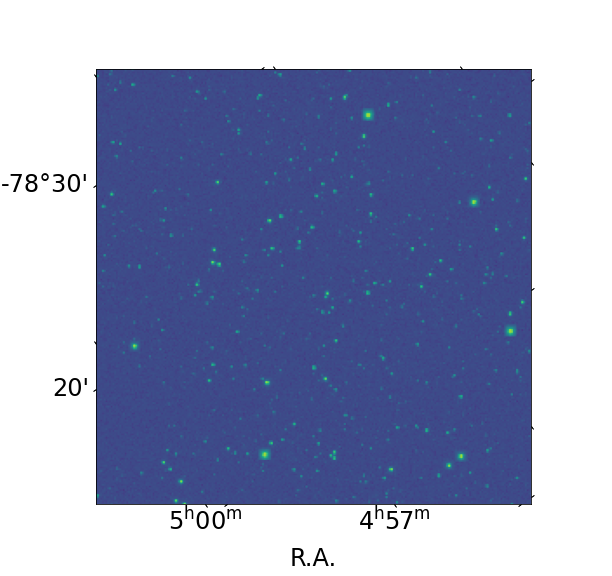} \\
   \includegraphics[height=8cm]{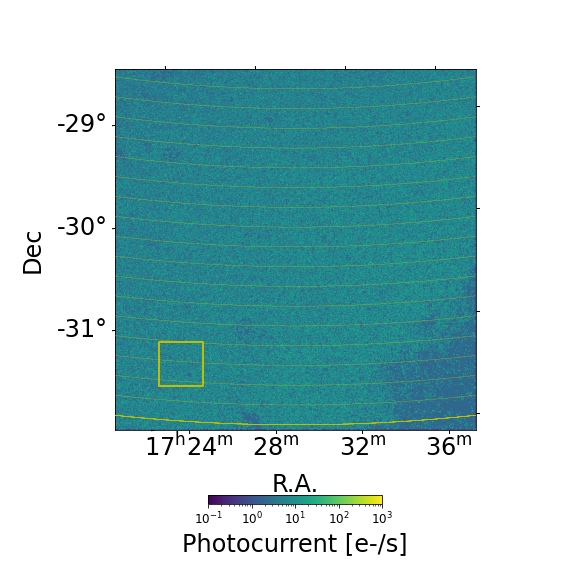} &
   \includegraphics[height=8cm]{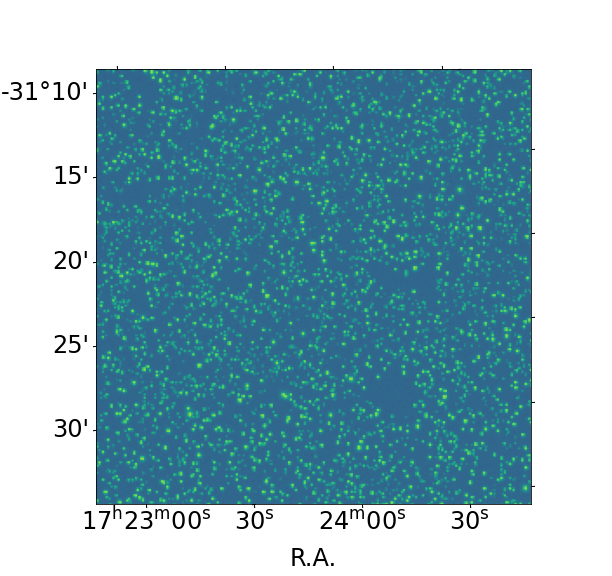} \\
   \end{tabular}
   \end{center}
   \caption[example] 
   { \label{fig:simulator} 
Simulated exposures of \spherex\ band 2 with the telescope pointing towards high ecliptic latitude (top row) and in the Galactic bulge (bottom row), including stars and zodiacal dust emission.  The left panels show the full 2048$\times$2048 array, with yellow contours indicating the pixel boundaries of the 17 spectral channels in this band, and the right panels show a zoomed-in portion of the array indicated with the yellow box. }
   \end{figure}

To model H2RG effects, an instrument model injects realizations of dark current and noise from the readout to each exposure.  In addition, models for full well/saturation effects,\cite{Biesiadzinski2011} subpixel response variations, and image persistence can be used optionally.  Realistic LVF performance, convolution of the simulated compact source signal with an arbitrary 2-dimensional PSF kernel is also applied, as well as optical field distortion from the telescope.  See Fig.~\ref{fig:simulator} for two example exposures.

The code is optimized to run in parallel in a High-Performance Computing environment, and can take advantage of GPU acceleration for PSF convolution.  Benchmarking on the Texas Advanced Computing Center\footnote{\url{http://www.tacc.utexas.edu}} LoneStar5 platform show that 6000 \spherex\ images (roughly two full days of data) can be generated using 6 core-hours, thus full-mission or full-survey-scale simulation runs are feasible.

\section{SPHEREX IN CONTEXT}
\begin{figure} [ht]
   \begin{center}
   \begin{tabular}{c} 
   \includegraphics[height=10cm]{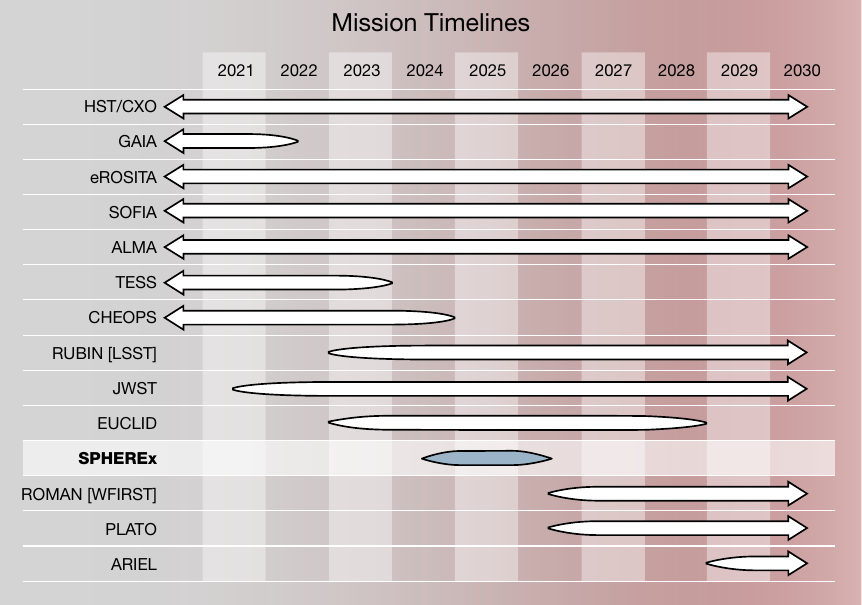}
   \end{tabular}
   \end{center}
   \caption[example] 
   { \label{fig:mission_timelines} 
Timeline for major missions and observatories which will be synergistic with \spherex\cite{Dore2018}. }
   \end{figure}

\begin{figure} [ht]
   \begin{center}
   \begin{tabular}{c} 
   \includegraphics[height=10cm]{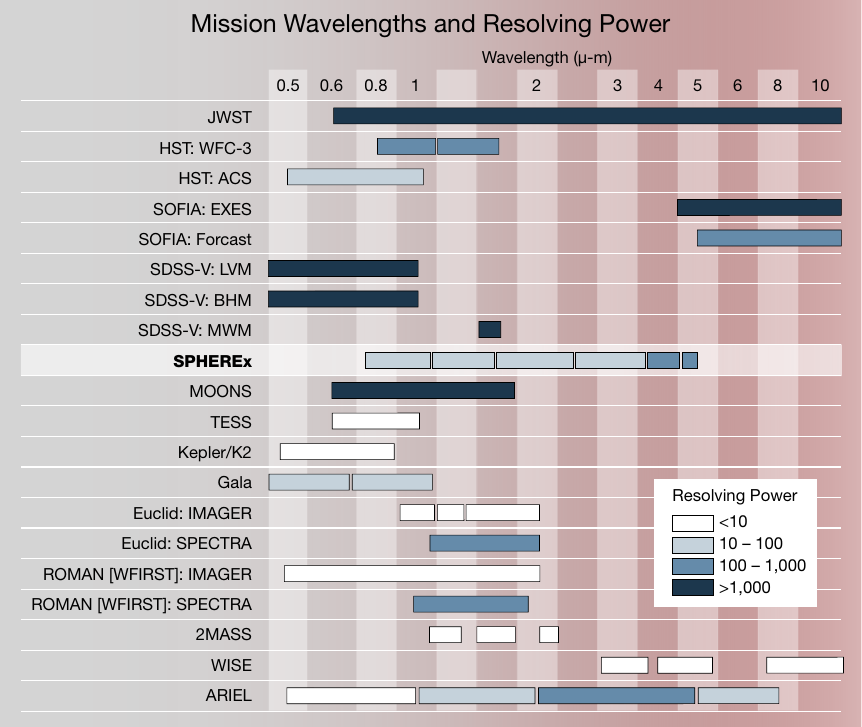}
   \end{tabular}
   \end{center}
   \caption[example] 
   { \label{fig:mission_wavelengths} 
Wavelength coverage and resolving power for major missions and observatories which will be synergistic with \spherex. }
   \end{figure}

The 2018 \spherex\ Community Workshop\cite{Dore2018}, focused on synergies between \spherex\  and other major astronomical facilities or missions operating now and into the 2020’s.  Such synergies are widespread and of great importance, enhancing the scientific return from both \spherex\  and the other facilities.  Figures~\ref{fig:mission_timelines} and~\ref{fig:mission_wavelengths} show \spherex\  temporal and measurement capability synergism with a number of these facilities and missions many of which, such as GAIA and eROSITA, are all sky surveys, while others, like Euclid, Roman, and Rubin, will survey large areas on the sky.  Thus the scientific coupling of these missions to \spherex’s all sky survey are particularly strong.
The timeline in Figure~\ref{fig:mission_timelines} shows that \spherex\ is well-timed to follow up on the results from missions such as JWST, TESS, and eROSITA; to identify targets for more detailed study by JWST, SOFIA, or ALMA, and to set the stage for later missions such as Roman (WFIRST) and PLATO.   Because \spherex\ is an all-sky survey itself, no advanced planning or target selection is required to optimize these synergies, which in some cases might not emerge until after both missions have been retired. 

The wavelength/spectral resolution chart shown in Figure~\ref{fig:mission_wavelengths} shows that \spherex\ provides almost unique access with significant spectral resolution to the wavelength range between 2.5 and 5\micron\ while extending shortward in wavelength to overlap with numerous ground-based and space-based imaging and spectroscopic surveys.  JWST is the only near-term space-based mission with comparable wavelength grasp and spectral resolution, and this makes the synergy between \spherex\ and JWST particularly strong, as \spherex\ goes wide while JWST goes deep.  At the end of the decade, the NASA/ESA mission ARIEL will also cover these wavelength though with targeted observations of exoplanet transits.  A number of spectroscopic surveys are planned for the next generation of operation of the Sloan Digital Sky Survey (SDSS), as well as the similar MOONS survey.  \spherex’s longer wavelength measurements, albeit at lower spectral resolution, will work well with these surveys, for example, by facilitating corrections for extinction or identifying circumstellar dust or cool companions not apparent at visible wavelengths.

Not shown in either chart are the 10--30-meter class telescopes currently operating on the ground and planned for the future.  Their overlap with \spherex\ will be similar to that of JWST, with the additional consideration that they will not have full access even to the 1-to-5\micron\ wavelength region because of atmospheric absorption.  Nevertheless, we can anticipate that these telescopes, with their much higher spectral and spatial resolution, will be effectively used in pursuing targets and scientific questions identified by \spherex.  The performance shown in Figure~\ref{fig:spherex_sensitivity} suggests that the \spherex\ spectra will achieve spectroscopy sensitivity within a factor of 2-to-3 of the broadband photometry sensitivity of the Gemini telescopes for long observations at wavelengths of three microns and beyond.

\section{CONCLUSION}
\spherex’s unprecedented all sky spectroscopic survey will open an important new spectroscopic window on the Universe, and we expect that the \spherex\ data will be used for scientific investigations going far beyond the three important science themes which define the mission.   The progress reported by Korngut et al (2018)\cite{Korngut2018} and in this paper shows that the \spherex\  design and technology are well in hand, leading to launch currently targeted for 2024.  \spherex\  will then become an important participant in a new era of astronomical exploration which has been called ``The Decade of the Surveys”.  In addition, we hope that the coupling of \spherex’s novel design to its driving scientific objectives may inspire similar innovation among groups developing modest-sized space missions in the future. 

\acknowledgments

Part of the research described in this paper was carried out at the Jet Propulsion Laboratory, California Institute of Technology, under a contract with the National Aeronautics and Space Administration (80NM0018D0004)

The authors acknowledge the Texas Advanced Computing Center (TACC) at the University of Texas at Austin for providing HPC resources that have contributed to the research results reported within this paper. 

This work has made use of data from the European Space Agency (ESA) mission Gaia (\url{https://www.cosmos.esa.int/gaia}), processed by the Gaia Data Processing and Analysis Consortium (DPAC, \url{https://www.cosmos.esa.int/web/gaia/dpac/consortium}). Funding for the DPAC has been provided by national institutions, in particular the institutions participating in the Gaia Multilateral Agreement.

This research made use of Astropy,\footnote{http://www.astropy.org} a community-developed core Python package for Astronomy \cite{astropy2013, astropy2018}. 

\bibliography{report} 
\bibliographystyle{spiebib} 

\end{document}